%% file: GramARIMA.tex
\input Definitions.tex
\centerline{\bf REGULAR BOHR-SOMMERFELD QUANTIZATION RULES} 

\centerline{\bf FOR A $h$-PSEUDO-DIFFERENTIAL OPERATOR~:}

\centerline{\bf THE METHOD OF POSITIVE COMMUTATORS}
\bigskip
\centerline{A.IFA ${}^{1}$ {\it\&} M. ROULEUX ${}^{2}$}
\bigskip
\centerline {${}^{1}$ Universit\'e de Tunis El-Manar, D\'epartement de Math\'ematiques, 1091 Tunis, Tunisia}

\centerline {e-mail: abdelwaheb.ifa@fsm.rnu.tn}

\centerline {${}^{2}$ Aix Marseille Univ, Univ Toulon, CNRS, CPT, Marseille, France} 

\centerline {e-mail: rouleux@univ-tln.fr}               
\bigskip
\noindent {\bf Abstract}: We revisit in this Note the well known Bohr-Sommerfeld quantization rule (BS) for 1-D Pseudo-differential 
self-adjoint Hamiltonians
within the algebraic and microlocal framework of Helffer and Sj\"ostrand; BS holds precisely when
the Gram matrix consisting of scalar products of WKB solutions with respect to the ``flux norm'' is not invertible. 
\medskip
\noindent{\bf R\'esum\'e}: Dans cette Note on propose une re-\'ecriture de la r\`egle de quantification de Bohr-Sommerfeld (BS)
pour un op\'erateur auto-adjoint $h$-Pseudo-diff\'erentiel 1-D, dans le cadre alg\'ebrique et microlocal \'elabor\'e par 
Helffer et Sj\"ostrand; elle est donn\'ee par la condition d'annulation du d\'eterminant de 
la matrice de Gram d'une paire de solutions WKB dans une base convenable, pour le produit scalaire associ\'e  
\`a la ``norme de flux''. 
\bigskip
\noindent {\bf 0. Introduction}.
\smallskip
Let $p(x,\xi;h)$ be a smooth real classical Hamiltonian on $T^*{\bf R}$, 
with the formal expansion $p\sim p_0(x,\xi)+hp_1(x,\xi)+h^2p_2(x,\xi)+\cdots$; we will assume that
$p$ belongs to the space of symbols $S^0(m)$ for some order function $m$ (for example $m(x,\xi)= (1+|\xi|^2)^M$) with
$$S^N(m)=\{p\in C^\infty(T^*{\bf R}): \forall\alpha\in{\bf N}^{2}, \exists C_\alpha>0,
|\partial^\alpha_{(x,\xi)}p(x,\xi;h)|\leq C_\alpha h^N m(x,\xi)\}$$
and has the semi-classical expansion
$p(x,\xi; h) \sim p_0(x,\xi)+hp_1(x,\xi)+\cdots, h\rightarrow 0$.
which allows to take Weyl quantization $P^w(x,hD_x;h)$ of $p$
$$P^w(x,hD_x)u(x;h)=(2\pi h)^{-1}\int\int e^{i(x-y)\eta/h} p({x+y\over2},\eta;h)u(y)\,dy\,d\eta\leqno(0.1)$$
so that $P^w(x,hD_x)$ is self-adjoint. We also assume that $p+i$ is elliptic (in the classical sense).
We call as usual $p_0$ the principal symbol, and $p_1$ the sub-principal symbol; in case of Schr\"odinger operator $P(x,hD_x)=(hD_x)^2+V(x)$,
$p(x,\xi;h)=p_0(x,\xi)=\xi^2+V(x)$. 
We make the hypothesis of [CdV], namely:

Fix some compact interval $I = [E_-,E_+], E_- < E_+$, and assume
that there exists a topological ring ${\cal A}\subset T^*{\bf R}$ such that $\partial{\cal A} = A_-\cup A_+$ with $A_\pm$ a
connected component of $p_0^{-1}(E_\pm)$. 
Assume also that $p_0$ has no critical point in ${\cal A}$, and
$A_-$ is included in the disk bounded by $A_+$ (if it is not the
case, we can always change $p$ to $-p$.)
We define the microlocal well $W$ as the disk bounded by $A_+$. 
For $E\in I$, let $\gamma_E\subset W$ be a periodic orbit in the energy surface $\{p_0(x,\xi)=E\}$ (so that $\gamma_E$ is an
embedded Lagrangian manifold). 

Let $K_h^N(E)$ be the microlocal kernel of $P-E$ of order $N$,
i.e. the space of local solutions of $(P-E)u={\cal O}(h^{N+1})$ in the distributional sense, microlocalized on $\gamma_E$. 
This is a smooth complex vector bundle over $\pi_x(\gamma_E)$. Here we address the problem of 
finding the set of $E=E(h)$ such that $K_h^N(E)$ contains a global section, i.e. of constructing quasi-modes (QM) $(u_n(h),E_n(h))$ 
of a given order $N$.

Then if $E_+<E_0=\liminf_{|x,\xi|\to\infty}p_0(x,\xi)$, all eigenvalues of $P$ in $I$ 
are indeed given by {\it Bohr-Sommerfeld quantization condition} (BS) $S_h(E_n(h))=2\pi nh$, where the {\it semi-classical action} $S_h(E)$ 
has the asymptotic
$S_h(E)\sim S_0(E)+hS_1(E)+h^2S_2(E)+\cdots$. We determine BS at any accuracy by computing quasi-modes.
There are a lot of ways to derive BS: the method of matching of WKB solutions [BenOrs], known also as Liouville-Green method [Ol],
which has received many improvements [Ya], 
the method of the monodromy operator (see [HeRo] and references therein), the method of quantization deformation
based on Functional Calculus and Trace Formulas [Li], [CdV], [CaGra-SazLiReiRios], [Gra-Saz], [Arg]. 
Note that the latter one already assumes BS, it only gives a very convenient way to derive it.
In the real analytic case, 
BS rule, and also tunneling expansions, can be obtained using the so-called ``exact WKB method'' see e.g. [FeMa], [DePh], [DeDiPh], [Ba] 
when $P=-h^2\Delta+V(x)$ is Schr\"odinger operator, or a system of such operators.  

Here we present another way to construct quasi-modes of order 2,
based on [Sj2], [HeSj]. We stress that our method in the present scalar 
case, when carried to second order, is a bit more intricated than [Li], [CdV] and its 
refinements [Gra-Saz]; it is most useful for matrix valued operators with double characteristics
such as Bogoliubov-de Gennes Hamiltonian ([BenIfaRo], [BenMhaRo]), or Born-Oppenheimer Hamiltonian [Ro]. 
\medskip
\noindent {\bf 1. The microlocal Wronskian}.
\smallskip
The best algebraic and microlocal framework for computing 1-D quantization rules in the self-adjoint case,
cast in the fundamental works [Sj2], [HeSj], is based on Fredholm theory,
and the classical ``positive commutator method'' using conservation of some quantity 
called a ``quantum flux''. 

Bohr-Sommerfeld quantization rules result in constructing quasi-modes by WKB approximation along a closed Lagrangian manifold
$\Lambda_E\subset\{p_0=E\}$, i.e. a periodic orbit of Hamilton vector field $H_p$ with energy $E$.
This can be done locally according to the rank of the projection $\Lambda_E\to{\bf R}_x$.

Thus asymptotic solutions to $(P-E)u=0$ along $\Lambda_E$ can be considered as (micro)-local sections of a bundle ${\cal M}_E$ over ${\bf R}$
with a compact base, corresponding to the ``classically allowed region'' at energy $E$. 
The sequence of eigenvalues
$E=E_n(h)$ is determined by the condition that the resulting quasi-mode, gluing together asymptotic
solutions from different coordinates patches along $\Lambda_E$,
be single-valued, i.e. ${\cal M}_E$ have trivial holonomy. 

Assuming $\Lambda_E$ is smoothly embedded in $T^*{\bf R}^2$, it can be always be parametrized by a non degenerate phase function. 
Of particular interest are the critical points of the phase functions, or {\it focal points}
which are responsible for the change in Maslov index. Recall that $a(E)=(x_E,\xi_E)\in\Lambda_E$ 
is a focal point if $\Lambda_E$ ``turns vertical'' at $a(E)$, i.e.
$T_{a(E)}\Lambda_E$ is no longer transverse to the fibers $x=\Const $ in $T^*{\bf R}$. 
In any case, however, $\Lambda_E$ can be parametrized locally either by a phase $S=S(x)$ (spatial representation) or a phase 
$S=\widetilde S(\xi)$
(Fourier representation). Choose an orientation on $\Lambda_E$ and for $a\in\Lambda_E$ (not necessarily a focal point), denote by
$\rho=\pm1$ its oriented segments near $a$. 
Let $\chi^a\in C_0^\infty({\bf R}^2)$ be a smooth cut-off equal to 1 near $a$, and $\omega^a_\rho$ a small neighborhood
of $\supp [P,\chi^a]\cap\Lambda_E$ near $\rho$.
Here $\chi^a$ holds for $\chi^a(x,hD_x)$ as in (0.1), and 
we shall equally write $P(x,hD_x)$ (spatial representation) or $P(-hD_\xi,\xi)$ (Fourier representation).
Let also $u^a,v^a\in K_h^N$ ($K_h$ for short) be asymptotic solutions of $(P-E)u=0$ supported on $\Lambda_E$. 
\medskip
\noindent{\bf Definition 1.1}: Let $P$ be self-adjoint. We call 
$${\cal W}^a_\rho(u^a,\overline{v^a})=\bigl({i\over h}[P,\chi^a]_\rho u^a|v^a\bigr)\leqno(1.1)$$ 
the {\it microlocal Wronskian} of $(u^a,\overline{v^a})$ in $\omega^a_\rho$. Here ${i\over h}[P,\chi^a]_\rho$ denotes the part of the 
commutator supported microlocally on $\omega^a_\rho$.  
\smallskip
To understand that terminology, let $P=-h^2\Delta+V$, $x_E=0$ and change $\chi$ to Heaviside unit step-function $\chi(x)$,
depending on $x$ alone. Then in distributional sense,
we have ${i\over h}[P,\chi]=-ih\delta'+2\delta hD_x$, where $\delta$ denotes the Dirac measure at 0, and $\delta'$ its derivative,
so that $\bigl({i\over h}[P,\chi]u|u\bigr)=-ih\bigl(u'(0)\overline{u(0)}-u(0)\overline{u'(0)}\bigr)$
is the usual Wronskian of $(u,\overline u)$.
\smallskip
\noindent{\bf Proposition 1.2}: Let $u^a,v^a\in K_h$ as above, and denote by $\widehat u$ the $h$-Fourier (unitary) transform of $u$. Then
$$\leqalignno{
&\bigl({i\over h}[P,\chi^a] u^a|v^a\bigr)=\bigl({i\over h}[P,\chi^a] \widehat u^a|\widehat v^a\bigr)=0&(1.2)\cr
&\bigl({i\over h}[P,\chi^a]_+ u^a|v^a\bigr)=-\bigl({i\over h}[P,\chi^a]_- u^a|v^a\bigr)&(1.3)\cr
}$$ 
(all equalities being understood mod ${\cal O}(h^\infty)$). Moreover, ${\cal W}^a_\rho(u^a,\overline{v^a})$ doesn't depend of the choice of $\chi^a$ as above.
\smallskip
\noindent {\it Proof}: Since $u^a,v^a\in K_h$ are tempered distributions, the first equality
(1.2) follows from Plancherel formula. If $a$ is not a focal point, $u^a,v^a$ are smooth WKB solutions near $a$, so we can
expand the commutator in ${W}=\bigl({i\over h}[P,\chi^a] u^a|v^a\bigr)$ and use that $P$ is self-adjoint to show that
${W}={\cal O}(h^\infty)$. If $a$ is a focal point, $u^a,v^a$ are smooth WKB solutions in Fourier representation, so again
${W}={\cal O}(h^\infty)$. Then (1.3) follows
from Definition 1.1. $\clubsuit$
\smallskip
We can find a linear combination of 
${\cal W}^a_\pm$, (depending on $a$) which defines a sesquilinear form on $K_h$, so that this Hermitean form makes 
of ${\cal M}_E$ a metric bundle, endowed with the gauge group $U(1)$. 
This linear combinaison is prescribed as the construction of Maslov index~: namely 
we take ${\cal W}^a(u^a,\overline{u^a})={\cal W}^a_+(u^a,\overline{u^a})-{\cal W}^a_-(u^a,\overline{u^a})>0$ when the critical 
point $a$ of $\pi_{\Lambda_E}$ is traversed in the $-\xi$ direction to the right of the fiber 
(or equivalently ${\cal W}^a(u^a,\overline{u^a})=-{\cal W}^a_+(u^a,\overline{u^a})+{\cal W}^a_-(u^a,\overline{u^a})>0$ 
when $a$ is traversed in the $+\xi$ direction to the left of the fiber). Otherwise, just exchange the signs. 
When $\gamma_E$ is a convex curve, there are only 2 focal points. In general there may be many focal points $a$, 
but each jump of Maslov index is compensated at the next focal point
which is traversed to the other side of the fiber (Maslov index is computed mod 4). 

Our method consists in constructing Gram matrix of a generating system of ${\cal M}_E$ in a suitable dual basis; 
its determinant vanishes precisely
at the eigenvalues $E(h)$. 
\medskip
\noindent{\bf 2. QM and BS in the case of a Schr\"odinger operator}.
\smallskip
As a warm-up, we derive the well known BS quantization rule using microlocal Wronskians
in case of a potential well, i.e. $\gamma_E$ is convex.
Consider the spectrum of Schr\"odinger operator 
$P(x,hD_x)=(hD_x)^2+V(x)$ near the energy level $E_0<\liminf_{|x|\to\infty}V(x)$, when $\{V\leq E\}=[x'_E,x_E]$
and $x'_E,x_E$ are simple turning points, $V(x'_E)=V(x_E)=E$, $V'(x'_E)<0,V'(x_E)>0$. 
It is convenient to start the construction from the focal points $a$ or $a'$.
We set $a'=x'_E$, $a=x_E$, identifying the focal point $a=a_E=(x_E,0)$ 
with its projection $x_E$. 
We know that microlocal solutions $u$ of $(P-E)u=0$ near $a$ are of the form
$$u^a(x,h)={C\over\sqrt2}\bigl(e^{i\pi/4}(E-V)^{-1/4}e^{iS(a,x)/h}+e^{-i\pi/4}(E-V)^{-1/4}e^{-iS(a,x)/h}+{\cal O}(h)\bigr), \ C\in{\bf C}\leqno(2.1)$$
where $S(y,x)=\int_y^x\xi_+(t)\, dt$, and $\xi_+(t)$ is the positive root of $\xi^2+V(t)=E$. 
In the same way, the microlocal solutions of  $(P-E)u=0$ near $a'$ have the form
$$u^{a'}(x,h)={C'\over\sqrt2}\bigl(e^{-i\pi/4}(E-V)^{-1/4}e^{iS(a',x)/h}+e^{i\pi/4}(E-V)^{-1/4}e^{-iS(a',x)/h}+{\cal O}(h)\bigr), \ C'\in{\bf C} 
\leqno(2.2)$$
These expressions result in computing by the method of stationary phase
the oscillatory integral that gives the solution of $(P(-hD_\xi,\xi)-E)\widehat u=0$
in Fourier representation. The change of phase factor $e^{\pm i\pi/4}$ accounts for Maslov index.
For the sake of simplicity, we omit henceforth ${\cal O}(h)$ terms, but the computations below
extend to all order in $h$ (practically, at least for $N=2$), thus giving the asymptotics of BS. This will be elaborated in Section 3.

The semi-classical distributions $u^a,u^{a'}$ span the microlocal kernel $K_h$ of $P-E$ in $(x,\xi)\in]a',a[\times{\bf R}$; they are 
normalized using microlocal Wronskians as follows.

Let $\chi^a\in C_0^\infty({\bf R}^2)$ as in the Introduction be a smooth cut-off equal to 1 near $a$.
Without loss of generality, we can take $\chi^a(x,\xi)=\chi^a_1(x)\chi_2(\xi)$, so that $\chi_2\equiv1$ on small neighborhoods $\omega^a_\pm$,
of $\supp [P,\chi^a]\cap\{\xi^2+V=E\}$ in $\pm\xi>0$. We define
$\chi^{a'}$ similarly.
By (2.1) and (2.2) we have, mod ${\cal O}(h)$:
$$\eqalign{
&{i\over h}[P,\chi^a]u^a(x,h)=
\sqrt2C(\chi^a_1)'(x)\bigl(e^{i\pi/4}(E-V)^{1/4}e^{iS(a,x)/h}-e^{-i\pi/4}(E-V)^{1/4}e^{-iS(a,x)/h}\bigr)\cr
&{i\over h}[P,\chi^{a'}]u^{a'}(x,h)=
\sqrt2C'(\chi^{a'}_1)'(x)\bigl(e^{-i\pi/4}(E-V)^{1/4}e^{iS(a',x)/h}-e^{i\pi/4}(E-V)^{1/4}e^{-iS(a',x)/h}\bigr)\cr
}$$
Let 
$$F^a_\pm(x,h)={i\over h}[P,\chi^a]_\pm u^a(x,h)=\pm\sqrt2C(\chi^a_1)'(x)e^{\pm i\pi/4}(E-V)^{1/4}e^{\pm iS(a,x)/h}$$ 
so that:
$$\eqalign{
&(u^a|F^a_+-F^a_-)=|C|^2\bigl(e^{i\pi/4}(E-V)^{-1/4}e^{iS(a,x)/h}|(\chi^a_1)'e^{i\pi/4}(E-V)^{1/4}e^{iS(a,x)/h})\cr
&+|C|^2(e^{-i\pi/4}(E-V)^{-1/4}e^{-iS(a,x)/h}|(\chi^a_1)'e^{-i\pi/4}(E-V)^{1/4}e^{-iS(a,x)/h})\bigr)+{\cal O}(h)\cr
&=|C|^2(\int(\chi^a_1)'(x)dx+\int(\chi^a_1)'(x)dx)+{\cal O}(h)=2|C|^2+{\cal O}(h)
}$$
(the mixed terms such as $\bigl(e^{i\pi/4}(E-V)^{-1/4}e^{iS(a,x)/h}|(\chi^a_1)'e^{-i\pi/4}(E-V)^{1/4}e^{-iS(a,x)/h})$ are ${\cal O}(h^\infty)$ because
the phase is non stationary), thus $u^a$ is normalized mod ${\cal O}(h)$
if we choose $C=2^{-1/2}$. 
In the same way, with 
$$F^{a'}_\pm(x,h)={i\over h}[P,\chi^{a'}]_\pm u^{a'}(x,h)=\pm\sqrt2C'(\chi^{a'}_1)'(x)e^{\mp i\pi/4}(E-V)^{1/4}e^{\pm iS(a',x)/h}$$ 
we get
$$(u^{a'}|F^{a'}_+-F^{a'}_-)=|C'|^2(\int(\chi^{a'}_1)'(x)dx+\int(\chi^{a'}_1)'(x)dx)+{\cal O}(h)=-2|C'|^2+{\cal O}(h)$$
and we choose again $C'=C$ which normalizes $u^{a'}$ mod ${\cal O}(h)$. Normalization carries to higher order, as is shown in Sect.3.
 
So there is a natural duality product between $K_h$ and the
span of functions $F^a_+-F^a_-$ and $F^{a'}_+-F^{a'}_-$ in $L^2$. As in [Sj2], [HeSj]
we can show that this space is microlocally transverse to $\im (P-E)$ on $(x,\xi)\in]a',a[\times{\bf R}$, and thus identifies with 
the microlocal co-kernel $K_h^*$ of
$P-E$; in general $\dim K_h=\dim K_h^*=2$, unless $E$ is an eigenvalue, in which case $\dim K_h=\dim K_h^*=1$ (showing that
$P-E$ is of index 0 when Fredholm.~)

Microlocal solutions $u^a$ and $u^{a'}$ extend as smooth solutions on the whole interval $]a',a[$; we denote them by $u_1$ and $u_2$.
Since there are no other focal points between $a$ 
and $a'$, they are expressed by the same formulae (which makes the analysis particularly simple) and satisfy~:
$$(u_1|F^{a}_+-F^{a}_-)=1,\quad (u_2|F^{a'}_+-F^{a'}_-)=-1$$
Next we compute (still modulo ${\cal O}(h)$)
$$\eqalign{
&(u_1|F^{a'}_+-F^{a'}_-)={1\over2}(e^{i\pi/4}(E-V)^{-1/4}e^{iS(a,x)/h}|(\chi^{a'}_1)'e^{-i\pi/4}(E-V)^{1/4}e^{iS(a',x)/h})\cr
&+{1\over2}(e^{-i\pi/4}(E-V)^{-1/4}e^{-iS(a,x)/h}|(\chi^{a'}_1)'e^{i\pi/4}(E-V)^{1/4}e^{-iS(a',x)/h})\cr
&={i\over2}e^{-iS(a',a)/h}\int(\chi^{a'}_1)'(x)dx-{i\over2}e^{iS(a',a)/h}\int(\chi^{a'}_1)'(x)dx=-\sin (S(a',a)/h)\cr
}$$
(taking again into account that the mixed terms are ${\cal O}(h^\infty)$). Similarly
$(u_2|F^{a}_+-F^{a}_-)=\sin (S(a',a)/h)$. Now we define Gram matrix
$$G^{(a,a')}(E)=\pmatrix{(u_1|F^{a}_+-F^{a}_-)&(u_2|F^{a}_+-F^{a}_-)\cr (u_1|F^{a'}_+-F^{a'}_-)&(u_2|F^{a'}_+-F^{a'}_-)\cr}\leqno(2.4)$$
whose determinant $-1+\sin^2 (S(a',a)/h)=-\cos^2 (S(a',a)/h)$ vanishes precisely on eigenvalues of $P$ in $I$, so we recover the 
well known BS quantization condition (mod ${\cal O}(h)$)
$$\oint\xi(x)\, dx=2\int_{a'}^a(E-V)^{1/2}\, dx=2\pi h(k+{1\over2})+{\cal O}(h)\leqno(2.5)$$
and $\det G^{(a,a')}(E)$ is nothing but Jost function which is computed e.g. in [DeDi], [DeDiPh] by another method. 
\medskip
\noindent{\bf 3. The general case}
\smallskip
By the discussion after Proposition 1.1, it is clear that it suffices to consider the case when $\gamma_E$ contains only 2 focal points
which contribute to Maslov index. 
\medskip
\noindent {\it a) Well normalized quasi-modes mod ${\cal O}(h^2)$}.
\smallskip
Let $a=(x_E,\xi_E)$ be such a focal point. Following a well known procedure we can trace back to [Sj1],
we first seek for asymptotic solutions in Fourier representation near $a$ of the form $\widehat u(\xi)=e^{i\psi(\xi)/h}b(\xi;h)$. 
Here the phase $\psi=\psi_E$ solves Hamilton-Jacobi equation $p_0(-\psi'(\xi),\xi)=E$, and can be normalized by $\psi(\xi_E)=0$;
the amplitude $b(\xi;h)=b_0(\xi)+hb_1(\xi)+\cdots$ has to be found recursively
together with $a(x,\xi;h)=a_0(x,\xi)+ha_1(x,\xi)+\cdots$, 
such that 
$$hD_\xi\bigl( e^{i(x\xi+\psi(\xi))/h} a(x,\xi;h)\bigr)=e^{i(x\xi+\psi(\xi))/h}b(\xi;h)\bigl(
p_0(x,\xi)-E+h\widetilde p_1(x,\xi)+h^2\widetilde p_2(x,\xi)+{\cal O}(h^3)\bigr)$$ 
$p_0$ being the principal symbol of $P$, $\widetilde p_1$ its sub-principal symbol for the standard (Feynman) quantization, etc\dots.
Define $\lambda(x,\xi)$ by $p_0(x,\xi)-E=\lambda(x,\xi)(x+\psi'(\xi))$, we get first
$$\lambda(-\psi'(\xi),\xi)=\partial_xp_0(-\psi'(\xi),\xi)=_{\fed }\alpha(\xi)$$
This yields $a_0(x,\xi)=\lambda(x,\xi)b_0(\xi)$ and solving a first order ODE $L(\xi,D_\xi)b_0=0$, with
$L(\xi,D_\xi)=\alpha(\xi)D_\xi+\bigl({1\over2i}\alpha'(\xi)-p_1(\-\psi'(\xi),\xi)\bigr)$ we get
$$b_0(\xi)=C_0|\alpha(\xi)|^{-1/2}e^{i\int{p_1\over\alpha}}$$
with an arbitrary constant $C_0$, we take independent of $E$. This gives in turn $a_1(x,\xi)=\lambda(x,\xi)b_1(\xi)+\lambda_0(x,\xi)$, with 
$$\lambda_0(x,\xi)={b_0(\xi)\widetilde p_1+i{\partial a_0\over\partial\xi}\over x+\partial_\xi\psi}$$
and $b_1(\xi)$ solution of $L(\xi,D_\xi)b_1=\widetilde p_2b_0+i\partial_\xi\lambda_0|_{x=-\psi'(\xi)}$. We eventually get
$$b_0(\xi)+hb_1(\xi)=(C_0+hC_1+hD_1(\xi))|\alpha(\xi)|^{-1/2}e^{i\int{p_1\over\alpha}}$$
where we have set
$$D_1(\xi)=\sgn\alpha(\xi_E)\int_{\xi_E}^\xi \bigl(i\widetilde p_2b_0-\partial_\xi\lambda_0|_{x=-\psi'(\xi')}\bigr)\, 
|\alpha(\xi')|^{-1/2}e^{-i\int{p_1\over\alpha}}\,d\xi'\leqno(3.1)$$
The integration constants $C_0,C_1,\cdots$ will be determined 
by normalizing the microlocal Wronskians as follows.

We compute ${\cal W}^a_\rho(u^a,\overline{u^a})$ in Fourier representation, with 
$\widehat u(\xi;h)=e^{i\psi(\xi)/h}b(\xi;h)$. Recall $\chi^a\in C_0^\infty({\bf R}^2)$, $\chi^a\equiv1$ near $a_E$; 
without loss of generality, we can take $\chi^a(x,\xi)=\chi_1(x)\chi_2(\xi)$, so that $\chi_2\equiv1$ on small neighborhoods $\omega^a_\pm$,
of $\supp {i\over h}[P,\chi^a]\cap\gamma_E$ in $\pm(\xi-\xi_E)>0$. Thus we need only consider the variations of $\chi_1$.
Weyl symbol of ${i\over h}[P,\chi^a]$ is given by  
$c(x,\xi;h)=\bigl(\partial_\xi p_0(x,\xi)+h\partial_\xi p_1(x,\xi)\bigr)\chi'_1(x)+{\cal O}(h^2)$, 
so
$${i\over h}[P,\chi^a]\widehat u(\xi)=(2\pi h)^{-1}\int\int 
e^{i\bigl(-(\xi-\eta)y+\psi(\eta)\bigr)/h}c(y,{\xi+\eta\over2};h)(b_0+hb_1)(\eta)\,dy\, d\eta$$
Evaluating by stationary phase, we find ${i\over h}[P,\chi^a]\widehat u(\xi)=e^{i\psi(\xi)/h}d(\xi;h)$, where 
$d(\xi;h)=d_0+hd_1+{\cal O}(h^2)$ with $d_0=c_0b_0$ and $d_1$ a function of $c_0,c_1,b_0,b_1$ we have determined so far. 
It follows 
$$\bigl({i\over h}[P,\chi^a]_+\widehat u|\widehat u\bigr)=\int_{\xi_E}^\infty d_0(\xi)\overline{b_0(\xi)}\,d\xi+{\cal O}(h)$$
The leading term is simply
$$\int_{\xi_E}^\infty|b_0|^2c_0\, d\xi=|C_0|^2\sgn(\alpha(\xi_E))\int_{\xi_E}^\infty \chi'_1(-\psi'(\xi))\psi''(\xi)\, d\xi
=|C_0|^2\sgn(\alpha(\xi_E))\leqno(3.2)$$
so $u^a$ is normalized mod ${\cal O}(h)$, provided $\alpha(\xi_E)>0$,
when $2|C_0|^2=1$; we take $C_0=1/\sqrt2$ as in Schr\"odinger case. 
Next step in normalization involves the term $D_1(\xi)$ defined in (3.1); integrating by parts, we can
remove $\chi_1(\xi)$ and its second derivative, so to end up with a simple integral like (3.2). 
The computation being rather lengthy, we only state the final result:
\medskip
\noindent {\bf Lemma 3.1}: With the hypotheses above, the microlocal Wronskian near a focal point $a_E$ is given by
$$\eqalign{
{\cal W}&^a(u^a,\overline{u^a})=\bigl({i\over h}([P,\chi^a]_+-[P,\chi^a]_-)\widehat u|\widehat u)=\cr
&2\sgn(\alpha(\xi_E))\bigl(|C_0|^2+h(2\re(\overline{C_0}C_1)+|C_0|^2\partial_x\bigl({p_1\over\partial_xp_0}\bigr)(\xi_E)+{\cal O}(h^2)\bigr)
\cr}$$
The condition that $u^a$ be normalized mod ${\cal O}(h^2)$ (once we have chosen $C_0$ to be real), is then
$$C_1=-{1\over2}C_0\partial_x\bigl({p_1\over\partial_xp_0}\bigr)(\xi_E)$$ 
so that now 
${\cal W}^a(u^a,\overline{u^a})=2C_0^2\bigl(1+{\cal O}(h^2)\bigr)$. This procedure carries to any order, we say then that
$u^a$ is {\it well-normalized}. It can be formalized by considering $\{a_E\}$ as a {\it Poincar\'e section}, and Poisson operator
the operator that assigns, 
in a 1-to-1 way, to the initial condition $C_0$ on $\{a_E\}$ the well-normalized (forward) solution $u^a$ to $(P-E)u^a=0$; see Sect.4.
\smallskip
The next task consists in extending the solutions away from $a_E$ in the spatial representation. First we 
expand $u^a(x)=(2\pi h)^{-1/2}\int e^{i\psi(\xi)/h}b(\xi;h)\,d\xi$ near $a$ by stationary phase, selecting the 2 critical points
$\xi_\rho(x)=\xi_\pm(x)$, that correspond to the phase functions $\varphi_\rho(x)=x\xi_\rho(x)+\psi(\xi_\rho(x))$. So we have
$$\eqalign{
u^{a}&(x;h)=u^{a}_{+}(x;h)+u^{a}_{-}(x;h)=\cr
&{1\over\sqrt2}\bigl[\bigl({\partial_\xi p_0(x,\xi_+(x))\over i}\bigr)^{-1/2}e^{iS_+(x_E,x;h)/h}+
\bigl({\partial_\xi p_0(x,\xi_-(x))\over i}\bigr)^{-1/2}e^{iS_-(x_E,x;h)/h}+{\cal O}(h)\bigr]\cr
}\leqno(3.3)
$$
where 
$$S_\pm(x_E,x;h)=\varphi_\pm(x_E,x)-h\int_{x_E}^{x}{p_1(y,\xi_\pm(y))\over\partial_\xi p_0(y,\xi_\pm(y)}\,dy+\sqrt2 
h^2\im\bigl(D_1(\xi_\pm(x))\bigr)\leqno(3.4)$$
and $\varphi_\pm(x_E,x)=x_E\xi_E+\int_{x_E}^x\xi_\pm(y)\, dy$. 
Then we use
standard WKB theory with Ansatz $u_\rho^a(x)=a_\rho(x;h)e^{i\varphi_\rho(x)/h}$. Omitting the index $\rho$, we find the usual half-density
$$a_0(x)={\widetilde C_0\over C_0}|\psi''(\xi(x))|^{-1/2}b_0(\xi(x))$$
with a new constant $\widetilde C_0\in{\bf R}$~; the next term is
$$a_1(x)=(\widetilde C_1+\widetilde D_1(x))|\beta_0(x)|^{-1/2}\exp\bigl(-i\int{p_1(x,\varphi'(x))\over\beta_0(x)}\,dx\bigr)$$
and $\widetilde D_1(x)$ a complex function with $\re \widetilde D_1(x)=-{1\over2}\widetilde C_0{\beta_1(x)\over\beta_0(x)}+\Const $,
$$\im \widetilde D_1(x)={\widetilde C_0}\bigl(\int{\beta_1(x)\over\beta_0^2(x)}p_1(x,\varphi'(x))\,dx
-\int{p_2(x,\varphi'(x))\over\beta_0(x)}\,dx\bigr)\leqno(3.5)$$
with $\beta_0(x)=\partial_\xi p_0(x,\varphi'(x))=-{\alpha(\xi(x))\over\xi'(x)}$, $\beta_1(x)=\partial_\xi p_1(x,\varphi'(x))$. 
\medskip
\noindent {\it b) The homology class of the generalized action}.
\smallskip
Here we identify the various terms in (3.1) and (3.5). 
First on $\gamma_E$ we have $\psi(\xi)=\int-x\, d\xi+\Const $, and $\varphi(x)=\int\xi\, dx+\Const $.  
By Hamilton equations 
$$\dot\xi(t)=-\partial_xp_0(x(t),\xi(t)), \quad \dot x(t)=\partial_\xi p_0(x(t),\xi(t))$$ 
so on $\gamma_E$ we have
$\int{p_1\over\alpha}\, d\xi=-\int{p_1\over\partial_\xi p_0}\, dx=-\int_{\gamma_E} p_1\, dt$. 
The form $p_1\, dt$ is called the subprincipal 1-form.
Next we consider $D_1(\xi)$ as the integral over $\gamma_E$ of the 1-form, defined near $a$ in Fourier representation as
$$\Omega_1=T_1\,d\xi=\sgn(\alpha(\xi))\bigl(i\widetilde p_2b_0-\partial_\xi\lambda_0\bigr)|\alpha|^{-1/2} e^{-i\int{p_1\over\alpha}}\, d\xi$$ 
Using WKB constructions, $\Omega_1$ can also be extended in the spatial representation. 
Since $\gamma_E$ is Lagrangian, $\Omega_1$ is a closed form that we are going to compute modulo exact forms. 
Using integration by parts, the integral of $\Omega_1(\xi)$ in Fourier representation simplifies to
$$\leqalignno{
&\sqrt2\re D_1(\zeta)=-{1\over2}\bigl[\partial_x\bigl({p_1\over\partial_xp_0}\bigr)]_{\xi_E}^\xi&(3.7)\cr
\sqrt2\im &D_1(\zeta)=\int_{\xi_E}^\xi T_1(\zeta)\,d\zeta+
\bigl[{\psi''\over6\alpha}\partial_x^3 p_0+{1\over4}\alpha'\partial_x^2 p_0\bigr]_{\xi_E}^\xi&(3.8)\cr
&T_1={1\over\alpha}
\bigl(p_2-{1\over8}\partial_x^2\partial_\xi^2p_0+{\psi''\over12}\partial_x^3\partial_\xi p_0+{(\psi'')^2\over24}\bigl(\partial_x^4 p_0
\bigr)\bigr)+{1\over8}{(\alpha')^2\over\alpha^3}
\partial_x^2 p_0+\cr
&{1\over6}\psi''{\alpha'\over\alpha^2}\partial_x^3 p_0
-{p_1\over\alpha^{2}}\bigl(\partial_x p_1-{p_1\over2\alpha}\partial_x^2 p_0\bigr)&(3.9)\cr
}$$
Eq. (3.7) already shows that $\re\Omega_1$ is exact.
We can carry the integration in $x$-variable between the focal points $a_E$ and $a'_E$, and in $\xi$-variable again
near $a'_E$. Now let
$\Omega(x,\xi)=f(x,\xi)\, dx+g(x,\xi)\, d\xi$, where $f(x,\xi), g(x,\xi)$ are any smooth functions on ${\cal A}$.  
By Stokes formula
$$\int_{\gamma_E}\Omega(x,\xi)=\int\int_{p_0\leq E}(\partial_xg-\partial_\xi f)\, dx\wedge d\xi$$
where, following [CdV], we have extend $p_0$ inside the disk bounded by $A_-$ 
so that it coincides with a harmonic oscillator in a neighborhood of the origin ($p_0(0)=0$, say).
Making the symplectic change of coordinates $(x,\xi)\mapsto(t,E)$:
$$\int\int_{p_0\leq E}(\partial_xg-\partial_\xi f)\, dx\wedge d\xi=\int_0^E\int_0^{T(E')}(\partial_xg-\partial_\xi f)\, dt\wedge dE'\leqno(3.10)$$
where $T(E')$ is the period of the flow of Hamilton vector field $H_{p_0}$ at energy $E'$ ($T(E')$ being a constant near 0). 
Using these expressions, we recover the well known action integrals (see e.g. [CdV]):
\medskip
\noindent {\bf Lemma 3.2}: Let $\Gamma\, dt$ be the restriction to $\gamma_E$ of the 1-form
$$\omega_0(x,\xi)=\bigl((\partial^2_xp_0)(\partial_\xi p_0)-(\partial_x\partial_\xi p_0)(\partial_x p_0)\bigr)\, dx+
\bigl((\partial_\xi p_0)(\partial_\xi\partial_x p_0)-(\partial^2_\xi p_0)(\partial_x p_0)\bigr)\, d\xi$$ 
We have $\re \oint_{\gamma_E}\Omega_1=0$, whereas
$$\im \oint_{\gamma_E}\Omega_1={1\over48}\bigl({d\over dE}\bigr)^2\oint_{\gamma_E}\Gamma\, dt-\oint_{\gamma_E}p_2\, dt-
{1\over2}{d\over dE}\oint_{\gamma_E}p_1^2\, dt\leqno(3.12)$$
\medskip
\noindent {\it c) BS quantization rule}.
\smallskip
Recall from (3.3) the asymptotic solution $u^{a}(x;h)$ near $a=a_E$. In the last term of (3.4) we can substitute (3.8) with $T_1$ as in (3.9),
that is, $\sqrt2\im\bigl(D_1(\xi_\pm(x))\bigr)=\int_{x_E}^{x} T_1(\xi_\pm(y))\xi'_\pm(y)\, dy$.
Similarly, the asymptotic solution near $a'=a'_E$ is given by
$$\eqalign{
u^{a'}&(x;h)=u^{a'}_{+}(x;h)+u^{a'}_{-}(x;h)=\cr
&{1\over\sqrt2}\bigl[e^{-i\pi/4}\bigl(\partial_\xi p_0(x,\xi_+(x))\bigr)^{-1/2}e^{iS_+(x'_E,x;h)/h}+
e^{i\pi/4}\bigl(\partial_\xi p_0(x,\xi_-(x))\bigr)^{-1/2}e^{iS_-(x'_E,x;h)/h}+{\cal O}(h)\bigr]\cr
}\leqno(3.13)
$$
where as in (3.3), using (3.8) and (3.9)
$$S_\rho(x'_E,x;h)=\varphi_\rho(x'_E,x)-h\int_{x'_E}^{x}{p_1(y,\xi_\rho(y))\over\partial_\xi p_0(y,\xi_\rho(y)}\,dy+
h^2\int_{x'_E}^{x} T_1(\xi_\rho(y))\xi'_\rho(y)\, dy\leqno(3.14)$$
and similarly for $S_\rho(x_E,x;h)$.
The semi-classical distributions $u^a,u^{a'}$ are well normalized as in Lemma 3.1. We compute 
$F^{a}_\rho(x,h)={i\over h}[P,\chi^{a}]_\rho u^{a}(x,h)$. Still mod ${\cal O}(h)$ 
$$F^{a}_\rho(x,h)={\rho\over\sqrt2}\bigl({\rho\partial_\xi p_0(x,\xi_\rho(x)\over i}\bigr)^{1/2}e^{iS_\rho(x_E,x;h)/h}(\chi_1^a)'(x)$$
and using that the mixed terms $(u^a_\pm|F^a_\mp)$ are ${\cal O}(h^\infty)$, we find $(u^a|F^a_+-F^a_-)\equiv 1$ mod ${\cal O}(h)$. 
In the same way, near $a'$ we have $(u^a|F^a_+-F^a_-)\equiv -1$. The normalized microlocal solutions $u^a$ and $u^{a'}$, 
uniquely extended along $\gamma_E$, are now called $u_1$ and $u_2$. They verify 
$$\eqalign{
&(u_1|F_+^{a'}-F_+^{a'})\equiv{i\over2}\bigl(e^{iA_-(x_E,x'_E;h)/h}-e^{iA_+(x_E,x'_E;h)/h}\bigr)\cr
&(u_1|F_+^{a}-F_+^{a})\equiv{i\over2}\bigl(e^{-iA_-(x_E,x'_E;h)/h}-e^{-iA_+(x_E,x'_E;h)/h}\bigr)\cr
}\leqno(3.15)$$
where the generalized actions are given by
$$\eqalign{
A&_\rho(x_E,x'_E;h)=S_\rho(x_E,x;h)-S_\rho(x'_E,x;h)=\cr
&(x_E-x'_E)\xi_E+\int_{x_E}^{x'_E}\xi_\rho(y)\, dy-h\int_{x_E}^{x'_E}
{p_1(y,\xi_+(y))\over\partial_\xi p_0(y,\xi_\rho(y)}\,dy
+h^2\int_{x_E}^{x'_E}T_1(\xi_\rho(y))\xi'_\rho(y)\,dy\cr
}\leqno(3.16)$$
We have 
$$\eqalign{
&\int_{x'_E}^{x_E}\bigl(\xi_+(y)-\xi_-(y)\bigr)\,dy=\oint_{\gamma_E}\xi(y)\,dy\cr
&\int_{x'_E}^{x_E}\bigl({p_1(y,\xi_+(y))\over\partial_\xi p_0(y,\xi_+(y))}-{p_1(y,\xi_-(y))\over\partial_\xi p_0(y,\xi_-(y))}\bigr)
\,dy=\int_{\gamma_E}p_1\,dt\cr
&\int_{x'_E}^{x_E}\bigl(T_1(\xi_+(y))\xi'_+(y)-T_1(\xi_-(y))\xi'_-(y)\bigr)\,dy=\im\oint_{\gamma_E}\Omega_1(\xi(y))\,dy\cr
}$$
On the other hand, Gram matrix as in (2.4)
has determinant 
$$-\cos^2 \bigl(A_-(x_E,x'_E;h)-A_+(x_E,x'_E;h)/2h)$$ 
which vanishes precisely when BS holds.
\medskip
\noindent {\bf 4. The discrete spectrum of $P$ in $I$}.
\smallskip
We adapt the argument of [SjZw]. 
It is convenient to think of $\{a_E\}$ and $\{a'_E\}$ as 
zero-dimensional ``Poincar\'e sections'' of 
$\gamma_E$.
Let $K^a(E)$ be the operator (Poisson operator) that assigns to its ``initial value'' 
$C_0\in L^2(\{a_E\})\approx{\bf R}$ the well
normalized solution $u(x;h)=\int e^{i(x\xi+\psi(\xi))/h}b(\xi;h)\, d\xi$
to $(P-E)u=0$ near $\{a_E\}$. By construction, we have:
$$\pm K^a(E)^*{i\over h}[P,\chi^a]_\pm K^a(E)=\Id_{a_E}=1\leqno(4.1)$$
We define objects ``connecting'' $a$ to $a'$ along $\gamma_E$ as follows: let $T=T(E)>0$ such that $\exp TH_{p_0}(a)=a'$.
Choose $\chi^a_f$ ($f$ for ``forward'') be
a cut-off function supported microlocally near $\gamma_E$, equal to 0 along
$\exp tH_{p_0}(a)$ for $t\leq \e $, equal to 1 along $\gamma_E$ for
$t\in[2\e ,T+\e ]$, and back to 0 next to $a'$, e.g. for $t\geq T+2\e $. 
Let similarly $\chi^a_b$ ($b$ for ``backward'') 
be a cut-off function supported microlocally near $\gamma_E$, equal to 1 along 
$\exp tH_{p_0}(a)$ for $t\in[-\e ,T-2\e ]$, and equal to 0 next to $a'$, e.g. for $t\geq T-\e $. 
By (4.1) we have 
$$\leqalignno{
&K^a(E)^*{i\over h}[P,\chi^a]_+K^a(E)=K^a(E)^*{i\over h}[P,\chi^a_f]K^a(E)=1&(4.3)\cr 
&-K^a(E)^*{i\over h}[P,\chi^a]_-K^a(E)=-K^a(E)^*{i\over h}[P,\chi^a_b]K^a(E)=1&(4.4)\cr
}$$ 
which define a left inverse $R_+^a(E)=K^a(E)^*{i\over h}[P,\chi^a_f]$ to $K^a(E)$ and a right inverse 
$$R_-^a(E)=-{i\over h}[P,\chi^a_b]K^a(E)$$ 
to $K^a(E)^*$.
We define similar objects connecting $a'$ to $a$, $T'=T'(E)>0$ such that $\exp T'H_{p_0}(a)=a'$ ($T=T'$
if $p_0$ is invariant by time reversal), in particular a
left inverse $R_+^{a'}(E)=K^{a'}(E)^*{i\over h}[P,\chi^{a'}_f]_+$ to $K^{a'}(E)$ and a right inverse 
$R_-^{a'}(E)=-{i\over h}[P,\chi^{a'}_b]K^{a'}(E)$ to $K^{a'}(E)^*$, with the additional requirement 
$$\chi^a_b+\chi^{a'}_b=1\leqno(4.5)$$ 
near $\gamma_E$. 
Define now the pair $R_+(E)u=(R^a_+(E)u,R^{a'}_+(E)u)$, $u\in L^2({\bf R})$ and $R_-(E)$ by $R_-(E)u_-=R^a_-(E)u^a_-+R^{a'}_-(E)u^{a'}_-$, 
$u_-=(u^{a}_-,u^{a'}_-)\in{\bf C}^2$, we call Grushin operator ${\cal P}(z)$ the operator defined by the linear system
$${i\over h}(P-z)u+R_-(z)u_-=v, \quad R_+(z)u=v_+\leqno(4.6)$$
From [SjZw], we know that the problem (4.6) is well posed, and ${\cal P}(z)^{-1}=\pmatrix{E(z)&E_+(z)\cr E_-(z)&E_{-+}(z)}$, with
$(P-z)^{-1}=E(z)-E_+(z)E_{-+}(z)^{-1}E_-(z)$. Actually one can show that the effective Hamiltonian $E_{-+}(z)$ is Gram matrix (2.4).
There follows that the spectrum of $P$ in $I$ is precisely the set of $E$ we have determined by BS quantization rule.
\medskip
\noindent {\bf References}
\smallskip
\noindent [Ar] P.Argyres. The Bohr-Sommerfeld quantization rule and Weyl correspondence, Physics 2, p.131-199 (1965)

\noindent [Ba] H.Baklouti. Asymptotique des largeurs de resonances pour un modele d'effet tunnel microlocal. Ann. Inst. H.Poincare (Phys.Th.)
68 2), p.179-228, 1998.

\noindent [BenOrs] C.Bender S.Orzsag. Advanced Mathematical Methods for Scientists and Engineers. Srpinger, 1979.

\noindent [BenIfaRo] A.Bensouissi, A.Ifa, M.Rouleux. Andreev reflection and the
semi-classical Bogoliubov-de Gennes Hamiltonian.
Proceedings ``Days of Diffraction 2009'', Saint-Petersburg. p.37-42. IEEE 2009.

\noindent [BenMhaRo] A.Bensouissi, N.M'hadbi, M.Rouleux. Andreev reflection and the
semi-classical Bogoliu- bov-de Gennes Hamiltonian: resonant states.
Proceedings ``Days of Diffraction 2011'', Saint-Peters- burg. p.39-44. IEEE 101109/DD.2011.6094362 

\noindent [CaGra-SazLittlReiRios] M.Cargo, A.Gracia-Saz, R.Littlejohn, M.Reinsch \& P.de Rios. 
Moyal star product approach to the Bohr-Sommerfeld approximation, J.Phys.A:Math and Gen.38, 1977-2004 (2005).

\noindent [CdV] Y.Colin de Verdi\`ere. Bohr Sommerfeld rules to all orders. Ann. H.Poincar\'e, 6, p.925-936, 2005.

\noindent [DePh] E.Delabaere, F.Pham. Resurgence methods in semi-classical asymptotics. Ann. Inst. H.Poin- car\'e 71(1), p.1-94, 1999.

\noindent [DeDiPh] E.Delabaere, H.Dillinger, F.Pham. Exact semi-classical expansions for 1-D quantum oscillators. J.Math.Phys. Vol.38 (12)
p.6126-6184 (1997)

\noindent [Gra-Saz] A.Gracia-Saz. The symbol of a function of a pseudo-differential operator. Ann. Inst. Fourier, 55(7), p.2257-2284 (2005)

\noindent [HeRo] B.Helffer, D.Robert. Puits de potentiel generalis\'{e}s et asymptotique semi-classique. Annales
Inst. H.Poincar\'{e} (Physique Th\'{e}orique), Vol.41, No 3, p.291-331 (1984)

\noindent [HeSj] B.Helffer, J.Sj\"ostrand. Semi-classical analysis for Harper's equation III. Memoire  No 39, Soc. Math. de France, 117 (4) 
(1988)

\noindent [Li] R.Littlejohn, Lie Algebraic Approach to Higher-Order Terms, Preprint June 2003.

\noindent [Ol] F.Olver Asymptotics and special functions. Academic Press, 1974.

\noindent [Ro] M.Rouleux. Tunneling effects for $h$-Pseudodifferential Operators, Feshbach Resonances and the 
Born-Oppenheimer Approximation {\it in}: Evolution Equations, Feshbach
Resonances, Singular Hodge Theory. Adv. Part. Diff. Eq. Wiley-VCH (1999)

\noindent [Sj] J.Sj\"ostrand. {\bf 1} Analytic singularities of solutions of boundary value problems. Proc. NATO ASI on 
Singularities in boundary value problems, D.Reidel, 1980, p.235-269.
{\bf 2} Density of states oscillations for magnetic Schrodinger operators, 
{\it in}: Bennewitz (ed.) Diff. Eq. Math. Phys. 1990. Univ. Alabama, Birmingham, p.295-345. 

\noindent [SjZw] J. Sj\"ostrand and M. Zworski. Quantum monodromy and semi-classical trace formulae, J. Math. Pure Appl.
81(2002), 1-33. 

\noindent [Vo] A.Voros. Asymptotic h-expansions of stationary quantum states, Ann. Inst. H. Poincar\'{e} Sect. A(N.S), 26 (1977), 343-403
\bye

%% file: Definitions.tex
\magnification=1100

\hsize 17truecm
\vsize 23truecm

\font\twelvec=msbm10 at 12pt
\font\sevenc=msbm10 at 9pt
\font\fivec=msbm10 at 7pt

\newfam\co
\textfont\co=\twelvec
\scriptfont\co=\sevenc
\scriptscriptfont\co=\fivec

\def\Const{\mathop{\rm Const.}\nolimits}

\def\det{\mathop{\rm det}\nolimits}

\def\exp{\mathop{\rm exp}\nolimits}

\def\Id{\mathop{\rm Id}\nolimits}
\def\im{\mathop{\rm Im}\nolimits}

\def\lim{\mathop{\rm lim}\nolimits}

\def\re{\mathop{\rm Re}\nolimits}
\def\supp{\mathop{\rm supp}\nolimits}

\def\sgn{\mathop{\rm sgn}\nolimits}

\def\fed{\mathop{\rm def}\nolimits}

\def\e{\mathop{\rm \varepsilon}\nolimits}

\baselineskip 15pt
